# Transfer-matrix approach to the problem of electrical conduction through a series of absorbers

Kamil Walczak

Institute of Physics, Adam Mickiewicz University
Umultowska 85, 61-614 Poznań, Poland

Here we study incoherent transport through molecular wire treated as a linear chain of absorbers, where the phase-breaking processes are modeled by the use of imaginary point-like potentials. The calculations are performed within a transfer matrix method of the scattering theory. An analytic expression for the transmission of a finite chain is obtained, while the electrical current is then computed with the help of the Tsu-Esaki formula. In particular, it is shown that the maximal current dependence on the wire length is exponential.

Recent advances in nanofabrication allow us to obtain molecular devices composed of long molecular wires [1,2] and individual DNA molecules [3-17] connected to metallic electrodes and to measure transport characteristics of such junctions. DNA itself could be suitable for charge conduction, because some of the pi-orbitals belonging to particular bases of DNA overlap quite well with each other along the axis of the molecular chain [18]. However, the actual magnitude of DNA conductivity and its physical mechanism is still under debate nowadays [19-22]. In particular, long DNA oligomers are expected to be true insulators, while short biomolecules are rather semiconductors with a relatively large HOMO-LUMO gap. Besides, the electrical conduction through longer molecular bridges is incoherent in the sense that the charge carrier is temporarily localized on molecular fragments, loosing its phase and exchanging energy with the molecule. The main purpose of this work is to study incoherent transport through molecular wire treated as a linear chain of $N$ absorbers. Here the phase-breaking processes are modeled by the use of imaginary (or optical) point-like potentials [23-27]. In this paper we obtain analytical formula for the transmission and discuss in detail features associated with $I(V)$ (current-voltage) characteristics. The presented model can be used to simulate incoherent transport through longer bridges or DNA molecules by the choice of appropriate parameters.

For completeness, first we consider one-dimensional transport process through a single point-like absorber. Let us assume that the Hamiltonian of the system is given through the relation:

$$H = -\frac{\hbar^2}{2m}\frac{d^2}{dx^2} + iV_0\delta(x), \qquad (1)$$

where: $m$ is an effective mass of the conduction electron, while $V_0$ is the so-called dephasing potential. Since we are interested in steady state solution, the time-independent Schrödinger equation is taken into account: $H\Psi(x) = E\Psi(x)$, which can be rewritten in the form:

$$\Psi''(x) = \left[2i\Omega\delta(x) - k^2\right]\Psi(x), \qquad (2)$$

where: $\Omega = mV_0/\hbar^2$ and $k = \sqrt{2mE}/\hbar$. Two-step integration procedure of Eq.2 in the infinitesimal neighborhood of the potential allows us to obtain the following boundary conditions:

$$\Psi_1'(0) - \Psi_2'(0) = 2i\Omega\Psi_2(0), \tag{3}$$

$$\Psi_1(0) = \Psi_2(0). \tag{4}$$

Here: $\Psi_1(x)$ is the electron wavefunction on the left hand side of the potential, while $\Psi_2(x)$ is the electron wavefunction on the right hand side of the potential. These conditions are related to the conservation of the current at the junction (Kirchoff's law is given through Eq.3) and the single valuedness of the wavefunction (Eq.4).

If we assume that the charge carrier is coming from the left electrode (source) – through the molecule – to the right electrode (drain), we can write down the following solutions of the Schrödinger equation in the two distinguished regions:

$$\Psi_1(x) = \exp(ikx) + r\exp(-ikx), \tag{5}$$

$$\Psi_2(x) = t\exp(ikx), \tag{6}$$

Imposing the boundary conditions at the point $x = 0$ we obtain the two simple equations:

$$1 + r = t, \tag{7}$$

$$1 - r = [1 + 2/\xi]t, \tag{8}$$

where: $\xi = k/\Omega$. From such expressions we can find the coefficients: $t = \xi/(\xi+1)$, $r = -1/(\xi+1)$. And therefore transmission $T = |t|^2$ and reflection $R = |r|^2$ functions can be easily calculated as follows:

$$T = \frac{\xi^2}{(1+\xi)^2}, \tag{9}$$

$$R = \frac{1}{(1+\xi)^2}. \tag{10}$$

Here we can find out that $T + R \neq 0$. It means that there is non-zero probability that the electron can be absorbed at the point $x = 0$ due to the imaginary potential. The absorption parameter $A = 1 - T - R$ can be expressed in the form:

$$A = \frac{2\xi}{(1+\xi)^2}. \tag{11}$$

In Fig.1 we plot all the mentioned probabilities versus the $\xi$-parameter. Here we can distinguish four characteristics cases: (i) for $\xi = 0$ we deal with the case of perfect reflection: $T = A = 0$ and $R = 1$, (ii) for $\xi = 1$ we deal with the case of maximal absorption: $T = R = 1/4$ and $A = 1/2$, (iii) for $\xi = 2$ coefficients for transmission and absorption achieve the same values: $T = A = 4/9$ and $R = 1/6$, (iv) for $\xi \to \infty$ we deal with the case of perfect transmission: $T = 1$ and $R = A = 0$.

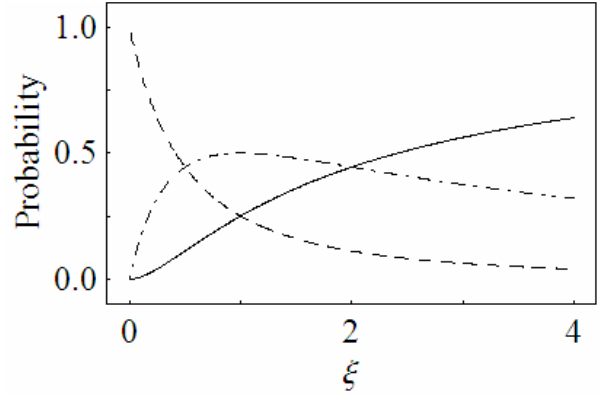

Figure 1: Transmission (solid line), reflection (dashed line) and absorption (dashed-dotted line) coefficients as functions of $\xi$ parameter for a single absorber.

Now let us consider a linear chain of $N$ point-like absorbers which are separated by the distance $L$. The molecule itself acts like a bridge, i.e. the potential drop across the molecule is very small. We can conceptually disassemble the system into the basic pieces connected in series [28]. The transfer matrix through one absorber is given through the relation:

$$M^{(m)} = \begin{bmatrix} 1/t^* & r/t \\ r^*/t^* & 1/t \end{bmatrix} = \begin{bmatrix} 1+1/\xi & -1/\xi \\ -1/\xi & 1+1/\xi \end{bmatrix}, \quad (12)$$

where: $m = 1,...,N$, $t$ and $r$ are the complex amplitudes of the transmission and reflection, while the asterisk (*) denotes the complex conjugation. The total transfer matrix for the case of $N$ point-like absorbers arranged in series can be expressed as:

$$M = M^{(N)} S M^{(N-1)} S ... M^{(2)} S M^{(1)}, \quad (13)$$

where:

$$S = \begin{bmatrix} \exp(ikL) & 0 \\ 0 & \exp(-ikL) \end{bmatrix}. \quad (14)$$

Assuming that all the absorbers are identical $M^{(N)} = M^{(N-1)} = ... = M^{(1)} = M_a$ and defining $M_a S \equiv \tilde{M}_a$, we get:

$$M = \left(\tilde{M}_a\right)^{N-1} M_a. \quad (15)$$

Matrix $\tilde{M}_a$ is not symmetric, but using similarity transformation it can be brought to symmetric form. This symmetric matrix can be diagonalized by an orthogonal matrix. Then the multiplication of $N$ matrices becomes very easy and we obtain the transfer matrix of the entire chain, which again has the form:

$$M = \begin{bmatrix} 1/t_N^* & r_N/t_N \\ r_N^*/t_N^* & 1/t_N \end{bmatrix}. \quad (16)$$

From this we can find the total transmission and reflection amplitudes across the periodic structure of $N$ absorbers and write them as:

$$t_N = \frac{2^{N+1}}{\eta} \sqrt{\varphi} \exp(-ikL), \tag{17}$$

$$r_N = \frac{2}{\eta}\left[\gamma_-^N - \gamma_+^N\right]. \tag{18}$$

Here we have introduced the following definitions:

$$\eta = \lambda_-\left[\chi_-^N - \chi_+^N\right] + \sqrt{\rho}\left[\chi_-^N + \chi_+^N\right], \tag{19}$$

$$\varphi = [1 + \exp(2ikL)]^2[\xi+1]^2 - 4\xi\exp(2ikL)[\xi+2], \tag{20}$$

$$\rho = 2\exp(2ikL)\left[1 - \xi(\xi+2) + (\xi+1)^2\cos(2kL)\right], \tag{21}$$

$$\chi_\pm = \frac{\exp(-ikL)}{\xi}\left[\lambda_+ \pm \sqrt{\rho}\right], \tag{22}$$

$$\gamma_\pm = \frac{\exp(-ikL)}{\xi}\left[\lambda_+ \pm \sqrt{\varphi}\right], \tag{23}$$

$$\lambda_\pm = [\xi+1][\exp(2ikL) \pm 1]. \tag{24}$$

The linear conductance of the entire network (at low voltages and in zero-temperature limit: $G = I/V$) is given through the Landauer relation as: $G = G_0 |t_N|^2$, where $G_0 = e^2/\pi\hbar \cong 77.5\,\mu S$ is the quantum of conductance.

The electrical current flowing through the junction can be calculated with the help of the Tsu-Esaki formula [29-32]:

$$I = I_0 \int\limits_{eV}^{+\infty} dE |t_N|^2 \ln\left(\frac{1 + \exp[\beta(E_F + eV - E)]}{1 + \exp[\beta(E_F - E)]}\right), \tag{25}$$

where:

$$I_0 = \frac{emB}{2\pi^2\hbar^3\beta}. \tag{26}$$

Here: $E_F$ is the Fermi energy, $V$ is the external voltage, $\beta = (k_B\theta)^{-1}$, $\theta$ is the device working temperature, $B \cong \pi r_s^2$ is the effective injection area of the transmitting electron from the metal electrode determined by the density of electronic states ($n$) of the bulk material, while $r_s = (3/4\pi n)^{1/3}$ is defined as the radius of a sphere whose volume is equal to the volume per conduction electron. The density of electronic states can be obtained by the following relation:

$$n = \frac{(2mE_F)^{3/2}}{3\pi^2\hbar^3}. \tag{27}$$

It should be noted that making use of the effective injection area conception we avoid the complications associated with the calculations of the self-energy [33].

To illustrate characteristic features of the linear conductance spectra we plot transmission against $kL$ for fixed $\xi$, see Fig.2. Here we can observe a well-defined band structure, where each allowed band for $N$ absorbers contains $N-1$ distinct peaks.

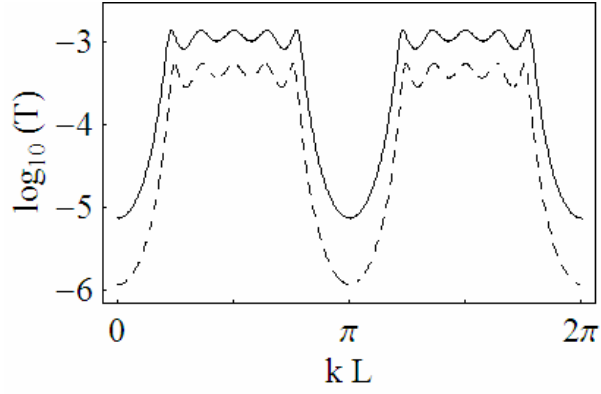

Figure 2: Transmission coefficients $T = |t_N|^2$ as functions of $kL$ for the case of $N = 6$ absorbers arranged in a linear chain for two different $\xi$-parameters: $\xi = 1.0$ (solid line) and $\xi = 0.8$ (dashed line). The picture is plotted in the logarithmic scale.

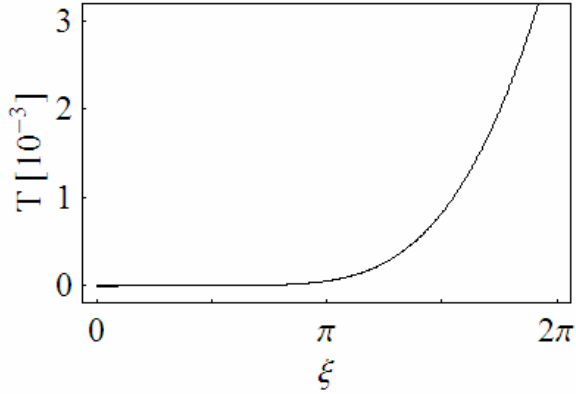

Figure 3: Transmission coefficient $T = |t_N|^2$ as functions of $\xi$-parameter for the case of $N = 20$ absorbers arranged in a linear chain for $kL = 1$.

Since the linewidth of each peak scales as $N^{-1}$, the conductance peak become sharper within a given band if the wire length is increased. Of course, the height of the peaks is reduced by orders of magnitude with respect to unity (perfect transmission), because here we have assumed large absorption. Decreasing the $\xi$-parameter (or increasing the dephasing $V_0$ potential), the transmission function does not change its shape but its values are reduced. In Fig.3 we present the transmission plotted against $\xi$ for fixed $kL$. Here we can see that for small values of $\xi (< \pi)$ – the transmission is close to zero, while for larger $\xi (> 1.5\pi)$ – transmission increases linearly with $\xi$-parameter. Interestingly, this picture does not change for a wide range of $kL$.

Now let us concentrate on the electrical current as a strongly nonlinear function of bias voltage and its length dependence. In Fig.4 we can see that the shape of the $I(V)$ curve is in qualitative agreement with measurements on DNA oligomers [4,5,11]. Our calculations indicate that the magnitude of the current flowing through the junction can change in a wide range of values from $\sim 50 \mu A$ for very short molecules ($\sim 1 nm$) to $\sim 25 pA$ for longer molecular wires ($\sim 50 nm$), as shown in Fig.5. For the model parameters chosen in this section, the following proportion is fulfilled:

$$\ln(I_{max}) \cong 11.76 - 0.34 \cdot N. \tag{28}$$

The magnitude of the maximal current given in $\mu A$ can be estimated from the relation:

$$I_{max} \cong 128 \times \exp(-0.34 \cdot N). \qquad (29)$$

Here we show that the maximal current dependence on the wire length is exponential: $I_{max} \propto \exp(-\alpha N)$, where $\alpha > 0$. Such conclusion is in contrast with recent theoretical calculations within rate-equation approach that indicate the algebraic length dependence: $I_{max} \propto N^{-\alpha}$, where $\alpha > 1$ for short molecules [34] and $\alpha = 1$ for very long biomolecules [35].

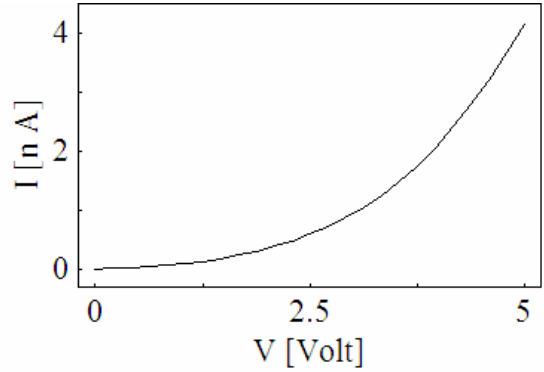

Figure 4: Electrical current as a bias voltage for $N = 30$ absorbers arranged in linear chain. Parameters: $m = 0.1 m_e = 9.11 \times 10^{-32} kg$, $V_0 = 5 \times 10^{-10} eV \cdot m$, $L = 5 \times 10^{-10} m$, $E_F = 5.3 eV$ (work function of the gold surface), $\theta = 300K$ ($\beta = 40/eV$).

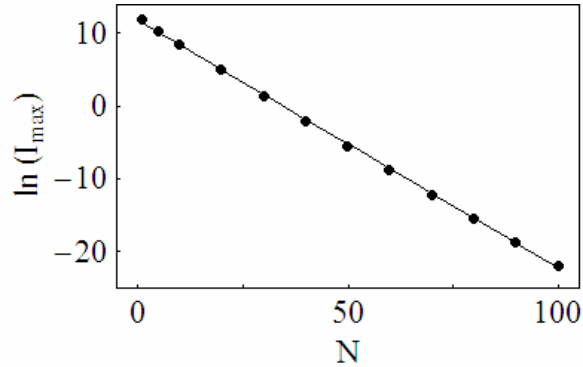

Figure 5: Maximal value of the current ($I_{max} = I(V = 5)$) flowing through the junction as a function of the number of absorbers ($N$). Model parameters are the same as in Fig.4. The picture is plotted in the logarithmic scale.

**Acknowledgement**

The author is very grateful to M. Sidowski and K. Herbeć for many interesting discussions.

**References**

[1] S.S. Isied, M.Y. Ogawa, and J.F. Wishart, Chem. Rev. **92**, 381 (1992).
[2] E.G. Petrov, Y.V. Schevchenko, V.I. Teslenko, and V. May,
    J. Chem. Phys. **115**, 7107 (2001).

[3] H.-W. Fink, and C. Schönenberger, Nature (London) **398**, 407 (1999).
[4] D. Porath, A. Bezryadin, S. de Vries, and C. Dekker, Nature (London) **403**, 635 (2000).
[5] L. Cai, H. Tabata, and T. Kawai, Appl. Phys. Lett. **77**, 3105 (2000).
[6] P. Tran, B. Alavi, and G. Gruner, Phys. Rev. Lett. **85**, 1564 (2000).
[7] A. Yu. Kasumov, M. Kociak, S. Gueron, B. Reulet, V.T. Volkov, D.V. Klinov, and H. Bouchiat, Science **291**, 280 (2001).
[8] A. Rakitin, P. Aich, C. Papadopoulos, Yu. Kobzar, A.S. Vedeneev, J.S. Lee, and J.M. Xu, Phys. Rev. Lett. **86**, 3670 (2001).
[9] K.-H. Yoo, D.H. Ha, J.-O. Lee, J.W. Park, J. Kim, J.J. Kim, H.-Y. Lee, T. Kawai, and H.Y. Choi, Phys. Rev. Lett. **87**, 198102 (2001).
[10] H. Watanabe, C. Manabe, T. Shigematsu, K. Shimotani, and M. Shimizu, Appl. Phys. Lett. **79**, 2462 (2001).
[11] J.S. Hwang, K.J. Kong, D. Ahn, G.S. Lee, D.J. Ahn, and S.W. Hwang, Appl. Phys. Lett. **81**, 1134 (2002).
[12] D.H. Ha, H. Nham, K.-H. Yoo, H. So, H.-Y. Lee, and T. Kawai, Chem. Phys. Lett. **355**, 405 (2002).
[13] B. Xu, P. Zhang, X. Li, and N. Tao, Nano Lett. **4**, 1105 (2004).
[14] B. Hartzell, B. McCord, D. Asare, H. Chen, J.J. Heremans, and V. Soghomonian, Appl. Phys. Lett. **82**, 4800 (2003).
[15] H. Cohen, C. Nogues, R. Naaman, and D. Porath, Proc. Natl. Acad. Sci. USA **102**, 11589 (2005).
[16] S. Tuukkanen, A. Kuzyk, J.J. Toppari, V.P. Hytönen, T. Ihalainen, and P. Törmä, Appl. Phys. Lett. **87**, 183102 (2005).
[17] R.M.M. Smeets, U.F. Keyser, D. Krapf, M.-Y. Wu, N.H. Dekker, and C. Dekker, Nano Lett. **6**, 89 (2006).
[18] D.D. Eley, and D.I. Spivey, Trans. Faraday Soc. **58**, 411 (1962).
[19] M.A. Ratner, Nature (London) **171**, 737 (1999).
[20] C. Dekker, and M.A. Ratner, Phys. World **14**, 29 (2001).
[21] R.G. Endres, D.L. Cox, and R.R.P. Singh, Rev. Mod. Phys. **76**, 195 (2004).
[22] M. Di Ventra, and M. Zwolak, *Encyclopedia of Nanoscience and Nanotechnology*, (Ed. H.S. Nalwa), Am. Sci. Publ. 2004, Vol.2, 475.
[23] Y. Zohta, and H. Ezawa, J. Appl. Phys. **72**, 3584 (1992).
[24] A. Rubio, and N. Kumar, Phys. Rev. B **47**, 2420 (1993).
[25] A.M. Jayannavar, Phys. Rev. B **49**, 14718 (1994).
[26] P.W. Brouwer, and C.W.J. Beenakker, Phys. Rev. B **55**, 4695 (1997).
[27] P.S. Deo, and A.M. Jayannavar, Phys. Rev. B **57**, 8809 (1998).
[28] P.S. Deo, and A.M. Jayannavar, Phys. Rev. B **50**, 11629 (1994).
[29] R. Tsu, and L. Esaki, Appl. Phys. Lett. **22**, 562 (1973).
[30] P.J. Turley, and S.W. Teitsworth, Phys. Rev. B **44**, 3199 (1991).
[31] Y. Luo, C.-K. Wang, and Y. Fu, J. Chem. Phys. **117**, 10283 (2002).
[32] Y. Luo, C.-K. Wang, and Y. Fu, Chem. Phys. Lett. **369**, 299 (2003).
[33] S. Datta, *Electronic Transport in Mesoscopic Systems*, Cambridge University Press, Cambridge 1995.
[34] K. Walczak, arXiv:cond-mat/0601380 (2006).
[35] M. Bixon, and J. Jortner, Chem. Phys. **319**, 273 (2005).